\documentclass[preprint,showpacs,byrevtex,groupedaddress,pra,twocolumn,10pt]{revtex4}
\usepackage{amsfonts}
\usepackage{amsmath}
\usepackage{amssymb}
\usepackage{graphicx}
\usepackage{dcolumn}
\usepackage{bm}

\begin{document}

\preprint{PRA/Manuscript}
\title[Density-functional calculations of inner-shell excitation]{%
Inner-shell excitation of open-shell atoms: A spin-dependent localized
Hartree-Fock density-functional calculation}
\author{Zhongyuan Zhou$^{1,2}$}
\author{Shih-I Chu$^{1}$}
\affiliation{$^{1}$Department of Chemistry, University of Kansas, Lawrence, KS 66045\\
$^{2}$Department of Physics and Astronomy, University of Kansas, Lawrence,
KS 66045}
\keywords{one two three}
\pacs{31.15.Ew, 32.80.Wr, 32.80.Rm}

\begin{abstract}
The spin-dependent localized Hartree-Fock (SLHF) density-functional approach
is extended to the treatment of the inner-shell excited-state calculation of
open-shell atomic systems. In this approach, the electron spin-orbitals in
an electronic configuration are obtained by solving Kohn-Sham (KS) equation
with SLHF exchange potential and the Slater's diagonal sum rule is used to
evaluate the multiplet energy of an inner-shell excited state from the
single-Slater-determinant energies of the electronic configurations
involved. This approach together with the correlation potentials and energy
functionals proposed by Perdew and Wang's (PW) or Lee, Yang, and Parr's
(LYP) have been used to calculate the total and excitation energies of
inner-shell excited states of open-shell atomic systems: Li, B, Ne$^{+}$, Ne$%
^{2+}$, Ne$^{3+}$, and Na. The results with the PW and LYP energy
functionals are in overall good agreement with each other and also with
available experimental and other \textit{ab initio} theoretical data. Some
new results for highly excited inner-shell states are presented.
\end{abstract}

\received{July 5, 2006}
\maketitle

\section{Introduction}

Due to computational simplicity and efficiency \cite{Parr89,Dreizler90}
density functional theory (DFT) \cite{Hohenberg64,Kohn65} has been widely
applied to many areas in theoretical physics and chemistry as a powerful 
\textit{ab initio} approach for the calculation of ground-state properties
of many-electron systems. The basic equation of DFT is Kohn-Sham (KS)
equation \cite{Kohn65} and the key part in KS equation is
exchange-correlation (XC) potential \cite{a16}. Due to incomplete
cancellation of spurious self-interactions and inherent degeneracy (due to
the use of spin and angular-momentum independent local potentials) of
traditional XC potentials obtained from uniform electron gas, such as local
density approximation (LDA) \cite{Parr89,Dreizler90} and generalized
gradient approximation (GGA) \cite{Parr89,Becke88,Perdew86,a38,a21}, the
conventional DFT using LDA or GGA is a ground-state approach. The
differences of the KS energy eigenvalues of unoccupied and occupied orbitals
are not rigorously defined as excitation energies. However, the KS energy
eigenvalues can serve as good zeroth-order excited-state energies provided
they are obtained by solving KS equation with a high-quality XC potential 
\cite{a15}. A number of theoretical methods have been developed by adopting
this point of view \cite{Singh99}. In particular, density work-functional
approach (WF) \cite{a11,a13,a14,a39}, open-shell localized Hartree-Fock
(LHF) density-functional approach \cite%
{Sala2003,Sala2003-1,Vitale05,Gorling05}, and multireference LHF
density-functional approach \cite{Hupp2003,Hupp2003-1}, etc., have been
successfully used to calculate excited-state properties of atomic and
molecular systems.

Recently, an exchange (X)-only LHF density-functional theory has been
proposed and successfully applied to ground-state calculations of atomic and
molecular systems \cite{a15}. In this X-only DFT, the exchange potential in
the KS equation is a LHF exchange potential derived under the assumption
that X-only KS determinant is equal to the Hartree-Fock (HF) determinant. We
have recently extended this approach to excited states of atomic and
molecular systems by assuming that the X-only KS determinant is also equal
to the HF determinant for excited states \cite{zhou2005}. Based on this
postulate we have developed a spin-dependent localized Hartree-Fock (SLHF)
density-functional approach for excited-state calculation of atomic and
molecular systems \cite{zhou2005}. In this approach, the exchange potential
in the KS equation is an exact nonvariational SLHF exchange potential
constructed for both the ground and excited states. The SLHF potential is an
analogue of the LHF potential. It is self-interaction free and exhibits the
correct long-range behavior. Further, the SLHF potential requires the use of
only the occupied orbitals and is dependent of the orbital symmetry of the
state. This approach associating with Slater's diagonal sum rule \cite%
{Slater60} has been successfully used to calculate multiply excited states
of valence electrons of atomic systems \cite{zhou2005} and inner-shell
excitation of close-shell atomic systems \cite{zhou06-inner-close-shell}
with accurate results.

In this paper, we extend the SLHF density-functional approach to inner-shell
excited states of open-shell atomic systems. We compute the total and
excitation energies of inner-shell excited states of open-shell atomic
systems: Li, B, Ne$^{+}$, Ne$^{2+}$, Ne$^{3+}$, and Na. In the calculation,
the correlation potentials and energy functionals proposed by Perdew and
Wang (PW) \cite{a21} and\ by Lee, Yang, and Parr (LYP) \cite{a38} are used
to estimate electron correlation effect. We will show that the calculated
results are in overall good agreement with available theoretical and
experimental data. We also present some new results for the highly excited
inner-shell states for the first time.

\section{Theoretical Method}

The SLHF density-functional approach has been discussed in Ref. \cite%
{zhou2005} in detail and is outlined in this section for convenience.

In spin-dependent density-functional approach, a spin-orbital $\varphi
_{i\sigma }\left( \mathbf{r}\right) $ of the $i$th electron with spin $%
\sigma $ ($\sigma =$ $\alpha $ and $\beta $ for spin-up and spin-down,
respectively) and its orbital energy $\varepsilon _{i\sigma }$ are
determined by the KS equation%
\begin{equation}
H_{\sigma }(\mathbf{r})\varphi _{i\sigma }\left( \mathbf{r}\right)
=\varepsilon _{i\sigma }\varphi _{i\sigma }\left( \mathbf{r}\right) ,
\label{e11}
\end{equation}%
where,%
\begin{equation}
H_{\sigma }(\mathbf{r})=-\frac{1}{2}\nabla ^{2}+V_{\sigma }^{\text{eff}%
}\left( \mathbf{r}\right) ,  \label{e11-1}
\end{equation}%
is the KS Hamiltonian and%
\begin{equation}
V_{\sigma }^{\text{eff}}\left( \mathbf{r}\right) =V_{ext}\left( \mathbf{r}%
\right) +V_{H}\left( \mathbf{r}\right) +V_{xc\sigma }\left( \mathbf{r}%
\right) ,  \label{e11-2}
\end{equation}%
is the local effective potential. In Eq. (\ref{e11-2}), $V_{ext}\left( 
\mathbf{r}\right) $ is the external potential, $V_{H}\left( \mathbf{r}%
\right) $ is Hartree potential (classical Coulomb electrostatic potential
between electrons), and $V_{xc\sigma }\left( \mathbf{r}\right) $ is the XC
potential.

For a given atomic system, the external potential $V_{ext}\left( \mathbf{r}%
\right) $ is known exactly. The Hartree potential $V_{H}\left( \mathbf{r}%
\right) $ is given by%
\begin{equation}
V_{H}\left( \mathbf{r}\right) =\int \frac{\rho \left( \mathbf{r}^{\prime
}\right) }{\left\vert \mathbf{r}-\mathbf{r}^{\prime }\right\vert }d\mathbf{r}%
^{\prime },  \label{e13}
\end{equation}%
where, $\rho \left( \mathbf{r}\right) =\rho _{\alpha }\left( \mathbf{r}%
\right) +\rho _{\beta }\left( \mathbf{r}\right) $ is the total electron
density and $\rho _{\sigma }\left( \mathbf{r}\right) $ (for $\sigma =\alpha $
and $\beta $) is the spin-dependent electron density defined by 
\begin{equation}
\rho _{\sigma }\left( \mathbf{r}\right) =\sum_{i=1}^{N_{\sigma }}w_{i\sigma
}\left\vert \varphi _{i\sigma }\left( \mathbf{r}\right) \right\vert ^{2}.
\label{e10}
\end{equation}%
Here $N_{\sigma }$ is the number of electrons with spin $\sigma $ and $%
w_{i\sigma }$ is the occupied number of electrons in the spin-orbital $%
\varphi _{i\sigma }\left( \mathbf{r}\right) $.

The XC potential can be decomposed into the exchange potential $V_{x\sigma
}\left( \mathbf{r}\right) $ and the correlation potential $V_{c\sigma
}\left( \mathbf{r}\right) $. In the SLHF density-functional approach, the
exchange potential is a SLHF exchange potential $V_{x\sigma }^{\text{SLHF}}(%
\mathbf{r})$. It is given by%
\begin{equation}
V_{x\sigma }^{\text{SLHF}}(\mathbf{r})=V_{x\sigma }^{\text{S}}(\mathbf{r}%
)+V_{x\sigma }^{\text{C}}(\mathbf{r}),  \label{e109}
\end{equation}%
where,%
\begin{equation}
V_{x\sigma }^{\text{S}}(\mathbf{r})=-\frac{1}{\rho _{\sigma }(\mathbf{r})}%
\sum_{i,j=1}^{N_{\sigma }}\gamma _{ij}^{\sigma }\left( \mathbf{r}\right)
\int \frac{\gamma _{ij}^{\sigma }\left( \mathbf{r}^{\prime }\right) }{%
\left\vert \mathbf{r}-\mathbf{r}^{\prime }\right\vert }d\mathbf{r}^{\prime },
\label{e110}
\end{equation}%
is the Slater potential \cite{Slater60} and%
\begin{equation}
V_{x\sigma }^{\text{C}}(\mathbf{r})=\frac{1}{\rho _{\sigma }(\mathbf{r})}%
\sum_{i,j=1}^{N_{\sigma }}\gamma _{ij}^{\sigma }\left( \mathbf{r}\right)
Q_{ij}^{\sigma },  \label{e111}
\end{equation}%
is a correction to Slater potential. In Eqs. (\ref{e110}) and (\ref{e111}) $%
\gamma _{ij}^{\sigma }\left( \mathbf{r}\right) $ and $Q_{ij}^{\sigma }$ are
defined by%
\begin{equation}
\gamma _{ij}^{\sigma }\left( \mathbf{r}\right) =\varphi _{i\sigma }(\mathbf{r%
})\varphi _{j\sigma }(\mathbf{r}),  \label{e111-1}
\end{equation}%
and%
\begin{equation}
Q_{ij}^{\sigma }=\left\langle \varphi _{j\sigma }\left\vert V_{x\sigma }^{%
\text{SLHF}}-V_{x\sigma }^{\text{NL}}\right\vert \varphi _{i\sigma
}\right\rangle ,  \label{e111-2}
\end{equation}%
where, $V_{x\sigma }^{\text{NL}}$ is a nonlocal exchange operator of the
form of HF exchange potential but constructed from KS spin-orbitals.

The SLHF exchange potential determined by Eqs. (\ref{e109})--(\ref{e111-2})
has two arbitrary additive constants. The physical orbitals can only be
obtained by the use of appropriate constants in the exchange potential \cite%
{a15}. To settle down the constants so as to pick up the physical orbitals,
it is required that the highest-occupied-orbital $N_{\sigma }$ of each spin $%
\sigma $ does not contribute to the correction term $V_{x\sigma }^{\text{C}}(%
\mathbf{r})$. In this case, the correction term $V_{x\sigma }^{\text{C}}(%
\mathbf{r})$ decays exponentially, the SLHF exchange potential behaves
asymptotically as Slater potential and thus approaches to $-1/r$ at long
range \cite{a15}.

In atomic systems, an electron spin-orbital is characterized by three
quantum numbers $n,$ $l,$ and $\sigma $, where $n$ and $l$ are the principal
quantum number and orbital angular momentum quantum number of the electron,
respectively. In the spherical coordinates, the spin-orbital $\varphi
_{i\sigma }\left( \mathbf{r}\right) $ of an electron with quantum numbers $%
n, $ $l,$ and $\sigma $ can be expressed by%
\begin{equation}
\varphi _{i\sigma }\left( \mathbf{r}\right) =\frac{R_{nl\sigma }(r)}{r}%
Y_{lm}(\theta ,\phi ),  \label{e204}
\end{equation}%
where, $R_{nl\sigma }(r)$ is the radial spin-orbital, $Y_{lm}(\theta ,\phi )$
is the spherical harmonic, $m$ is the azimuthal quantum number, and $i$ is a
set of quantum numbers apart from spin $\sigma $ of the spin-orbital. The
radial spin-orbital $R_{nl\sigma }(r)$ is governed by radial KS equation,%
\begin{equation}
\left[ -\frac{1}{2}\frac{d^{2}}{dr^{2}}+\frac{l(l+1)}{2r^{2}}+v_{\sigma }^{%
\text{eff}}(r)\right] R_{nl\sigma }=\varepsilon _{nl\sigma }R_{nl\sigma },
\label{e205}
\end{equation}%
where $v_{\sigma }^{\text{eff}}(r)$ is the radial effective potential given
by 
\begin{equation}
v_{\sigma }^{\text{eff}}(r)=v_{ext}\left( r\right) +v_{H}\left( r\right)
+v_{x\sigma }^{\text{SLHF}}\left( r\right) +v_{c\sigma }\left( r\right) .
\label{e206-1}
\end{equation}%
In Eq. (\ref{e206-1}), $v_{ext}\left( r\right) $, $v_{H}\left( r\right) $, $%
v_{x\sigma }^{\text{SLHF}}\left( r\right) $, and $v_{c\sigma }\left(
r\right) $ are the radial external potential, radial Hartree potential,
radial SLHF exchange potential, and radial correlation potential,
respectively.

For an atomic system with a nuclear charge $Z$, the external potential is
the Coulomb potential between electron and nucleus%
\begin{equation}
v_{ext}\left( r\right) =-\frac{Z}{r}.  \label{e207}
\end{equation}

In central-field approach, the radial Hartree potential is calculated from%
\begin{equation}
v_{H}\left( r\right) =4\pi \int \frac{1}{r_{>}}\rho (r^{\prime })r^{\prime
2}dr^{\prime },  \label{e210}
\end{equation}%
where, $r_{>}$ is the larger of $r$ and $r^{\prime }$, $\rho (r)=\rho
_{\alpha }(r)+\rho _{\beta }(r)$ is the spherically averaged total electron
density, and $\rho _{\sigma }(r)$ $\left( \sigma =\alpha \text{ or }\beta
\right) $ is the spherically averaged spin-dependent electron density given
by%
\begin{equation}
\rho _{\sigma }(r)=\frac{1}{4\pi }\int \rho _{\sigma }(\mathbf{r})d\Omega =%
\frac{1}{4\pi }\sum_{nl}^{\nu _{\sigma }}w_{nl\sigma }\left[ \frac{%
R_{nl\sigma }}{r}\right] ^{2}.  \label{e204-1}
\end{equation}%
Here the symbol $\nu _{\sigma }$ stands for a set of quantum numbers for
summation and the sum is performed over all the occupied spin-orbitals with
spin $\sigma $. This expression is accurate for spherically symmetric
(close-shell) states, but it is only an approximation for non-spherically
symmetric (open-shell) states. It may induce an error when it is used to
evaluate the energy of a non-spherically symmetric state. However, the error
is negligible compared to the order of calculated multiplet splitting \cite%
{Singh99}.

The radial SLHF exchange potential is given by%
\begin{equation}
v_{x\sigma }^{\text{SLHF}}\left( r\right) =v_{x\sigma }^{\text{S}}\left(
r\right) +v_{x\sigma }^{\text{C}}\left( r\right) ,  \label{e211}
\end{equation}%
where,%
\begin{equation}
v_{x\sigma }^{\text{S}}\left( r\right) =-\frac{1}{4\pi \rho _{\sigma }(r)}%
\sum_{nlm}^{\nu _{\sigma }}\sum_{n^{\prime }l^{\prime }m^{\prime }}^{\nu
_{\sigma }}s_{nlm,n^{\prime }l^{\prime }m^{\prime }}^{\sigma }(r),
\label{e212-1}
\end{equation}%
is the radial Slater potential and%
\begin{equation}
v_{x\sigma }^{\text{C}}\left( r\right) =\frac{1}{4\pi \rho _{\sigma }(r)}%
\sum_{nlm}^{\nu _{\sigma }}\sum_{n^{\prime }l^{\prime }m^{\prime }}^{\nu
_{\sigma }}c_{nlm,n^{\prime }l^{\prime }m^{\prime }}^{\sigma }(r).
\label{e213}
\end{equation}%
is a correction to the radial Slater potential. The matrix elements $%
s_{nlm,n^{\prime }l^{\prime }m^{\prime }}^{\sigma }(r)$ and $%
c_{nlm,n^{\prime }l^{\prime }m^{\prime }}^{\sigma }(r)$ in Eq. (\ref{e213})
are given in Ref. \cite{zhou2005}.

To calculate electron spin-orbital, the Legendre generalized pseudospectral
(LGPS) method \cite{Wang94} is used to discretize the radial KS equation (%
\ref{e205}). This method associated with an appropriate mapping technique
can overcome difficulties due to singularity at $r=0$ and long-tail at large 
$r$ of Coulomb interaction and thus provides a very effective and efficient
numerical algorithm for high-precision solution of KS equation. Using the
electron spin-orbitals of an electronic configuration, a single Slater
determinant for a specific electronic state is constructed and its total
energy calculated. The total energy is a sum of non-interacting
kinetic-energy $E_{k}$, external-field energy $E_{ext}$, Hartree energy $%
E_{H}$, exchange energy $E_{x}$, and correlation energy $E_{c}$. The values
of $E_{k}$, $E_{ext}$, $E_{H}$, and $E_{x}$ are evaluated by%
\begin{eqnarray}
E_{k} &=&\sum_{\sigma =\alpha }^{\beta }\sum_{nl}^{\nu _{\sigma
}}w_{nl\sigma }\int R_{nl\sigma }\left( r\right) \left( -\frac{1}{2}\frac{%
d^{2}}{dr^{2}}\right.  \notag \\
&&+\left. \frac{l(l+1)}{2r^{2}}\right) R_{nl\sigma }\left( r\right) dr,
\label{e216}
\end{eqnarray}%
\begin{equation}
E_{ext}=4\pi \int v_{ext}\left( r\right) \rho \left( r\right) r^{2}dr,
\label{e217}
\end{equation}%
\begin{equation}
E_{H}=\frac{1}{2}\sum\limits_{\Pi }\eta _{lm,l^{\prime }m^{\prime
}}^{k}F_{nl\sigma ,n^{\prime }l^{\prime }\sigma ^{\prime }}^{k},
\label{e218}
\end{equation}%
and%
\begin{equation}
E_{x}=-\frac{1}{2}\sum\limits_{\Pi }\lambda _{lm,l^{\prime }m^{\prime
}}^{k}G_{nl\sigma ,n^{\prime }l^{\prime }\sigma ^{\prime }}^{k}\delta
_{\sigma \sigma ^{\prime }},  \label{e221}
\end{equation}%
where, $\Pi $ represents a collection of all the quantum numbers involved,
the matrix elements $\eta _{lm,l^{\prime }m^{\prime }}^{k}$, $F_{nl\sigma
,n^{\prime }l^{\prime }\sigma ^{\prime }}^{k}$, and $G_{nl\sigma ,n^{\prime
}l^{\prime }\sigma ^{\prime }}^{k}$ are given in Ref. \cite{zhou2005}.

For a multiplet state that can be described completely by a single Slater
determinant, the energy is calculated directly from the single Slater
determinant. For a multiplet state that cannot be represented by a single
determinant, the energy can be calculated by means of Slater's diagonal sum
rule \cite{Slater60}. According to this rule, a sum over
single-Slater-determinant energy $E($D$_{i})$ of determinant D$_{i}$ from an
electron configuration equals to a weighted sum over multiplet energy $E($M$%
_{j})$ of multiplet state M$_{j}$ involved in the same electron
configuration, namely,%
\begin{equation}
\sum_{i}E(\text{D}_{i})=\sum_{j}d_{j}E(\text{M}_{j}),  \label{e223}
\end{equation}%
where, the weight $d_{j}$ is the times that the multiplet state M$_{j}$
appears in all the single Slater determinants. Similar procedures have been
employed in recent excited-state calculations \cite{a39,a14,Pollak97}.

\section{Results and discussion}

The procedure described in the preceding section is extended to calculate
the total energies $\left( E\right) $ and excitation energies $\left( \Delta
E\right) $ of inner-shell excited states of open-shell atomic systems: Li,
B, Ne$^{+}$, Ne$^{2+}$, Ne$^{3+}$, and Na. In the calculations, the
correlation effect, which is characterized by the correlation potential $%
v_{c\sigma }\left( r\right) $ and correlation energy $E_{c}$, is taken into
account through the correlation potentials and energy functionals of Perdew
and Wang (PW) \cite{a21} and of Lee, Yang, and Parr (LYP) \cite{a38},
respectively. The results obtained with these two correlation energy
functionals are listed in columns PW and LYP in the following tables,
respectively.

\subsection{Inner-shell excitation of Li}

Inner-shell excitation of Li has been the subject of extensive experimental
and theoretical studies. Both experimental \cite%
{Feldman67,Ederer70,Ziem75,Pegg75,Rassi77,McIlrath77,Cantu77,Rodbro79,Mannervik85}
and theoretical data \cite%
{Roy97b,Davis84,Davis88,Davis89,Chen94,Wakid80,Bhatia76,Bunge78,Bunge78-1,Bunge79,Bunge81,Lunell77-0,Lunell77,Holoien67}
are abundant. In TABLE \ref{T-1} to TABLE \ref{T-3} we present the total and
excitation energies of our calculations with PW and LYP correlation
potentials and energy functionals for inner-shell excited states $1s2sns$ $%
^{2,4}S$ $(n=2\thicksim 8)$, $1s2snp$ $^{2,4}P$ $(n=2\thicksim 8)$, and $%
1s2p^{2}$ $^{2}S$, $^{2,4}P$, and $^{2}D$, respectively. For comparison we
also list in these tables representative experimental data \cite%
{Ziem75,Rassi77,Rodbro79} as well as theoretical results obtained from
calculations using density work-functional formalism (WF) \cite{Roy97b},
saddle-point complex-rotation (SPCR) \cite{Davis84,Davis88,Davis89,Chen94},
extensive configuration-interaction calculation (ECI) \cite%
{Bunge78,Bunge78-1,Bunge79,Bunge81}, and Rayleigh-Ritz variational method
(RRV) \cite{Holoien67}.

For the total energies of inner-shell excited states $1s2sns$ $^{2,4}S$
listed in TABLE \ref{T-1}, the maximum relative discrepancies of our PW and
LYP results are 0.10\% and 0.32\% to the WF results, 0.21\%\ and 0.43\% to
the SPCR results, and 0.19\% and 0.62\% to the ECI results, respectively.
This indicates that both the PW and LYP results are in good agreement with
WF, SPCR, and ECI results, and the PW results are a little bit better than
the LYP results. For the excitation energy of excited state $1s2s^{2}$ $%
^{2}S $, the maximum relative deviations of our PW and LYP results and the
WF results are 0.91\%, 1.65\%, and 1.25\%, respectively, to the experimental
results. This demonstrates that the PW result is better than both the LYP
and WF results. For excited states $1s2sns$ $^{2}S$ $(n\geq 3)$, other
theoretical results of excitation energy are not available. For these states
the maximum relative discrepancies of our PW and LYP results of excitation
energies to the experimental results are 1.45\% and 2.75\%, respectively.
For the excited state $1s2s3s$ $^{4}S$, the maximum relative deviations of
our PW and LYP results of excitation energies to the experimental result are
0.75\% and 1.60\%, respectively. While the maximum relative discrepancies of
the ECI and RRV results to the experimental result are 0.59\% and 0.50\%,
respectively. Hence our PW result is in good agreement with both the
experimental result and other theoretical results for this state. For
excited states $1s2sns$ $^{4}S$ $(n>3)$ there is no experimental data
available. For these states, the maximum relative deviations of our PW and
LYP results to the RRV results are 0.68\% and 1.39\%, respectively. Thus our
PW results are close to the RRV results. For the all the results above the
PW results are better than the LYP results. As will be shown, due to
overestimation of LYP energy functional to correlation energies of atomic
systems with smaller $Z$ \cite{a38,zhou2005}, the LYP results are generally
worse than the PW results for atomic systems with $Z\leq 8$.

For the total energies of excited states $1s2snp$ $^{2,4}P$ given in TABLE %
\ref{T-2}, the maximum relative discrepancies of our PW and LYP results are
0.56\% \ and 0.89\% to the WF results, 0.64\% and 1.03\% to the SPCR
results, and 0.25\% and 0.82\% to the ECI results, respectively. This
illustrates that both the PW and LYP results are in agreement with the
complicated \textit{ab initio} methods. For the excitation energy of excited
state $1s2s2p$ $^{4}P$, the maximum relative deviations of our PW and LYP
results to the experimental results are 0.38\% and 1.66\%, while the maximum
relative discrepancies of the WF, ECI, and RRV results to the experimental
results are 1.423\%, 0.03\%, and 0.07\%, respectively. For excitation energy
of the excited state $1s2s2p$ $^{2}P$, the maximum relative discrepancies of
the PW, LYP, and WF results to the experimental result are 0.05\%, 1.70\%,
and 1.85\%, respectively. For excitation energy of the state $1s2s3p$ $^{2}P$%
, the maximum relative deviations of the PW, LYP, and WF results to the
experimental result are 1.35\%, 2.73\%, and 1.11\%, respectively. Thus the
PW results are better than the LYP results and a little bit better than the
WF results.

For the total energies of inner-shell doubly excited states $1s2p^{2}$ $%
^{2}S $, $^{2,4}P$, and $^{2}D$ given in TABLE \ref{T-3}, the maximum
relative deviations of our PW and LYP results are 0.63\% and 0.75\% to the
WF results, 0.47\% and 0.78\% to both the SPCR and ECI results,
respectively. This demonstrates that the PW energy functional has almost the
same precision as the LYP energy functional in calculation of total energies
of these inner-shell excited states. For excitation energies, the relative
discrepancies of our PW and LYP results to the experimental result are less
than 1.10\% and 1.50\%, respectively, while the maximum relative
discrepancies of the WF results to the experimental results are 1.09\%.
Hence the PW results are very close to the WF results and a little bit
better than the LYP results.

\begingroup\squeezetable

\begin{table*}[htbp] \centering%
\caption{Total energies ($E$) and excitation energies ($\Delta E$) of inner-shell 
excited states $1s2sns$ $^{2,4}S$ ($n=2 \thicksim 8$) of Li. 
The ground state energies obtained from calculations with PW and LYP correlation 
potentials and energy functionals are $-7.4837$ (a.u.) and $-7.4872$ (a.u.), 
respectively. Here 1 a.u.=27.2116 eV is used.\label{T-1}}%
\begin{tabular}{ccccccccccccccccccc}
\hline\hline
& \ \  &  & $-E$ &  &  & (a.u.) &  & \ \  &  &  &  & $\Delta E$ &  &  &  & 
(eV) &  &  \\ \cline{3-8}\cline{4-8}\cline{10-19}
States &  & Present & work &  & Other &  & theory &  & Present & work &  & 
Other &  & theory &  &  & Exp. &  \\ 
\cline{3-4}\cline{6-8}\cline{10-11}\cline{13-15}\cline{17-19}
&  & PW$^{a}$ & LYP$^{b}$ &  & WF$^{c}$ & SPCR$^{d}$ & ECI$^{e}$ &  & PW$%
^{a} $ & LYP$^{b}$ &  & WF$^{c}$ & ECI$^{e}$ & RRV$^{f}$ &  & ZBS$^{g}$ & RPR%
$^{h} $ & RBB$^{i}$ \\ \hline
\multicolumn{1}{r}{$1s2s^{2}$ $^{2}S$} &  & \multicolumn{1}{l}{$5.3940$} & 
\multicolumn{1}{l}{$5.3822$} & \multicolumn{1}{l}{} & \multicolumn{1}{l}{
5.3994} & \multicolumn{1}{l}{5.4052} & \multicolumn{1}{l}{} & 
\multicolumn{1}{l}{} & \multicolumn{1}{l}{$56.8641$} & \multicolumn{1}{l}{$%
57.2796$} & \multicolumn{1}{l}{} & \multicolumn{1}{l}{57.055} & 
\multicolumn{1}{l}{} & \multicolumn{1}{l}{} & \multicolumn{1}{l}{} & 
\multicolumn{1}{l}{56.352} & \multicolumn{1}{l}{56.395} & \multicolumn{1}{l}{
56.362} \\ 
\multicolumn{1}{r}{$1s2s3s$ $^{4}S$} &  & \multicolumn{1}{l}{$5.2225$} & 
\multicolumn{1}{l}{$5.2451$} & \multicolumn{1}{l}{} & \multicolumn{1}{l}{} & 
\multicolumn{1}{l}{} & \multicolumn{1}{l}{5.2127} & \multicolumn{1}{l}{} & 
\multicolumn{1}{l}{$61.5317$} & \multicolumn{1}{l}{$61.0114$} & 
\multicolumn{1}{l}{} & \multicolumn{1}{l}{} & \multicolumn{1}{l}{61.637} & 
\multicolumn{1}{l}{61.69} & \multicolumn{1}{l}{} & \multicolumn{1}{l}{} & 
\multicolumn{1}{l}{62.00} & \multicolumn{1}{l}{} \\ 
\multicolumn{1}{r}{$^{2}S$} &  & \multicolumn{1}{l}{$5.2019^{\star }$} & 
\multicolumn{1}{l}{$5.1466^{\star }$} & \multicolumn{1}{l}{} & 
\multicolumn{1}{l}{} & \multicolumn{1}{l}{} & \multicolumn{1}{l}{} & 
\multicolumn{1}{l}{} & \multicolumn{1}{l}{$62.0917^{\star }$} & 
\multicolumn{1}{l}{$63.6907^{\star }$} & \multicolumn{1}{l}{} & 
\multicolumn{1}{l}{} & \multicolumn{1}{l}{} & \multicolumn{1}{l}{} & 
\multicolumn{1}{l}{} & \multicolumn{1}{l}{61.995} & \multicolumn{1}{l}{} & 
\multicolumn{1}{l}{62.012$^{\star }$} \\ 
\multicolumn{1}{r}{} &  & \multicolumn{1}{l}{$5.1475^{\dagger }$} & 
\multicolumn{1}{l}{$5.1418^{\dagger }$} & \multicolumn{1}{l}{} & 
\multicolumn{1}{l}{} & \multicolumn{1}{l}{} & \multicolumn{1}{l}{} & 
\multicolumn{1}{l}{} & \multicolumn{1}{l}{$63.5728^{\dagger }$} & 
\multicolumn{1}{l}{$63.8310^{\dagger }$} & \multicolumn{1}{l}{} & 
\multicolumn{1}{l}{} & \multicolumn{1}{l}{} & \multicolumn{1}{l}{} & 
\multicolumn{1}{l}{} & \multicolumn{1}{l}{} & \multicolumn{1}{l}{} & 
\multicolumn{1}{l}{63.192$^{\dagger }$} \\ 
\multicolumn{1}{r}{$1s2s4s$ $^{4}S$} &  & \multicolumn{1}{l}{$5.1703$} & 
\multicolumn{1}{l}{$5.1918$} & \multicolumn{1}{l}{} & \multicolumn{1}{l}{} & 
\multicolumn{1}{l}{} & \multicolumn{1}{l}{} & \multicolumn{1}{l}{} & 
\multicolumn{1}{l}{$62.9513$} & \multicolumn{1}{l}{$62.4615$} & 
\multicolumn{1}{l}{} & \multicolumn{1}{l}{} & \multicolumn{1}{l}{63.113} & 
\multicolumn{1}{l}{63.18} & \multicolumn{1}{l}{} & \multicolumn{1}{l}{} & 
\multicolumn{1}{l}{} & \multicolumn{1}{l}{} \\ 
\multicolumn{1}{r}{$^{2}S$} &  & \multicolumn{1}{l}{$5.1288$} & 
\multicolumn{1}{l}{$5.1022$} & \multicolumn{1}{l}{} & \multicolumn{1}{l}{} & 
\multicolumn{1}{l}{} & \multicolumn{1}{l}{} & \multicolumn{1}{l}{} & 
\multicolumn{1}{l}{$64.0806$} & \multicolumn{1}{l}{$64.8999$} & 
\multicolumn{1}{l}{} & \multicolumn{1}{l}{} & \multicolumn{1}{l}{} & 
\multicolumn{1}{l}{} & \multicolumn{1}{l}{} & \multicolumn{1}{l}{63.17} & 
\multicolumn{1}{l}{63.16} & \multicolumn{1}{l}{63.292} \\ 
\multicolumn{1}{r}{$1s2s5s$ $^{4}S$} &  & \multicolumn{1}{l}{$5.1508$} & 
\multicolumn{1}{l}{$5.1715$} & \multicolumn{1}{l}{} & \multicolumn{1}{l}{} & 
\multicolumn{1}{l}{} & \multicolumn{1}{l}{} & \multicolumn{1}{l}{} & 
\multicolumn{1}{l}{$63.4819$} & \multicolumn{1}{l}{$63.0142$} & 
\multicolumn{1}{l}{} & \multicolumn{1}{l}{} & \multicolumn{1}{l}{} & 
\multicolumn{1}{l}{63.73} & \multicolumn{1}{l}{} & \multicolumn{1}{l}{} & 
\multicolumn{1}{l}{} & \multicolumn{1}{l}{} \\ 
\multicolumn{1}{r}{$^{2}S$} &  & \multicolumn{1}{l}{$5.1108$} & 
\multicolumn{1}{l}{$5.0855$} & \multicolumn{1}{l}{} & \multicolumn{1}{l}{} & 
\multicolumn{1}{l}{} & \multicolumn{1}{l}{} & \multicolumn{1}{l}{} & 
\multicolumn{1}{l}{$64.5704$} & \multicolumn{1}{l}{$65.3549$} & 
\multicolumn{1}{l}{} & \multicolumn{1}{l}{} & \multicolumn{1}{l}{} & 
\multicolumn{1}{l}{} & \multicolumn{1}{l}{} & \multicolumn{1}{l}{} & 
\multicolumn{1}{l}{} & \multicolumn{1}{l}{63.792} \\ 
\multicolumn{1}{r}{$1s2s6s$ $^{4}S$} &  & \multicolumn{1}{l}{$5.1412$} & 
\multicolumn{1}{l}{$5.1616$} & \multicolumn{1}{l}{} & \multicolumn{1}{l}{} & 
\multicolumn{1}{l}{} & \multicolumn{1}{l}{} & \multicolumn{1}{l}{} & 
\multicolumn{1}{l}{$63.7432$} & \multicolumn{1}{l}{$63.2847$} & 
\multicolumn{1}{l}{} & \multicolumn{1}{l}{} & \multicolumn{1}{l}{} & 
\multicolumn{1}{l}{63.99} & \multicolumn{1}{l}{} & \multicolumn{1}{l}{} & 
\multicolumn{1}{l}{} & \multicolumn{1}{l}{} \\ 
\multicolumn{1}{r}{$^{2}S$} &  & \multicolumn{1}{l}{$5.1018$} & 
\multicolumn{1}{l}{$5.0772$} & \multicolumn{1}{l}{} & \multicolumn{1}{l}{} & 
\multicolumn{1}{l}{} & \multicolumn{1}{l}{} & \multicolumn{1}{l}{} & 
\multicolumn{1}{l}{$64.8153$} & \multicolumn{1}{l}{$65.5810$} & 
\multicolumn{1}{l}{} & \multicolumn{1}{l}{} & \multicolumn{1}{l}{} & 
\multicolumn{1}{l}{} & \multicolumn{1}{l}{} & \multicolumn{1}{l}{} & 
\multicolumn{1}{l}{} & \multicolumn{1}{l}{} \\ 
\multicolumn{1}{r}{$1s2s7s$ $^{4}S$} &  & \multicolumn{1}{l}{$5.1358$} & 
\multicolumn{1}{l}{$5.1560$} & \multicolumn{1}{l}{} & \multicolumn{1}{l}{} & 
\multicolumn{1}{l}{} & \multicolumn{1}{l}{} & \multicolumn{1}{l}{} & 
\multicolumn{1}{l}{$63.8901$} & \multicolumn{1}{l}{$63.4370$} & 
\multicolumn{1}{l}{} & \multicolumn{1}{l}{} & \multicolumn{1}{l}{} & 
\multicolumn{1}{l}{64.33} & \multicolumn{1}{l}{} & \multicolumn{1}{l}{} & 
\multicolumn{1}{l}{} & \multicolumn{1}{l}{} \\ 
\multicolumn{1}{r}{$^{2}S$} &  & \multicolumn{1}{l}{$5.0966$} & 
\multicolumn{1}{l}{$5.0724$} & \multicolumn{1}{l}{} & \multicolumn{1}{l}{} & 
\multicolumn{1}{l}{} & \multicolumn{1}{l}{} & \multicolumn{1}{l}{} & 
\multicolumn{1}{l}{$64.9568$} & \multicolumn{1}{l}{$65.7100$} & 
\multicolumn{1}{l}{} & \multicolumn{1}{l}{} & \multicolumn{1}{l}{} & 
\multicolumn{1}{l}{} & \multicolumn{1}{l}{} & \multicolumn{1}{l}{} & 
\multicolumn{1}{l}{} & \multicolumn{1}{l}{} \\ 
\multicolumn{1}{r}{$1s2s8s$ $^{4}S$} &  & \multicolumn{1}{l}{$5.1324$} & 
\multicolumn{1}{l}{$5.1525$} & \multicolumn{1}{l}{} & \multicolumn{1}{l}{} & 
\multicolumn{1}{l}{} & \multicolumn{1}{l}{} & \multicolumn{1}{l}{} & 
\multicolumn{1}{l}{$63.9826$} & \multicolumn{1}{l}{$63.5315$} & 
\multicolumn{1}{l}{} & \multicolumn{1}{l}{} & \multicolumn{1}{l}{} & 
\multicolumn{1}{l}{} & \multicolumn{1}{l}{} & \multicolumn{1}{l}{} & 
\multicolumn{1}{l}{} & \multicolumn{1}{l}{} \\ 
\multicolumn{1}{r}{$^{2}S$} &  & \multicolumn{1}{l}{$5.0933$} & 
\multicolumn{1}{l}{$5.0695$} & \multicolumn{1}{l}{} & \multicolumn{1}{l}{} & 
\multicolumn{1}{l}{} & \multicolumn{1}{l}{} & \multicolumn{1}{l}{} & 
\multicolumn{1}{l}{$65.0466$} & \multicolumn{1}{l}{$65.7909$} & 
\multicolumn{1}{l}{} & \multicolumn{1}{l}{} & \multicolumn{1}{l}{} & 
\multicolumn{1}{l}{} & \multicolumn{1}{l}{} & \multicolumn{1}{l}{} & 
\multicolumn{1}{l}{} & \multicolumn{1}{l}{} \\ \hline\hline
\end{tabular}

$^{a}$PW results, $^{b}$LYP results, $^{c}$\cite{Roy97b}, $^{d}$\cite%
{Davis84}, $^{e}$\cite{Bunge78,Bunge78-1}, $^{f}$\cite{Holoien67}, $^{g}$%
\cite{Ziem75}, $^{h}$\cite{Rassi77}, and $^{i}$\cite{Rodbro79}. Here, $%
^{\star }$for $1s2s(^{3}S)3s$ $^{2}S$ and $^{\dagger }$for $1s2s(^{1}S)3s$ $%
^{2}S$.%
\end{table*}%
\endgroup

\begingroup\squeezetable

\begin{table*}[htbp] \centering%
\caption{Total energies ($E$) and excitation energies ($\Delta E$) of inner-shell
excited states $1s2snp$ $^{2,4}P$ ($n=2 \thicksim 8$) of Li.\label{T-2}}%
\begin{tabular}{ccccccccccccccccccc}
\hline\hline
& \ \  &  & $-E$ &  &  & (a.u.) &  & \ \  &  &  &  & $\Delta E$ &  &  &  & 
(eV) &  &  \\ \cline{3-8}\cline{10-19}
States &  & Present & work &  & Other &  & theory &  & Present & work &  & 
Other &  & theory &  &  & Exp. &  \\ 
\cline{3-4}\cline{6-8}\cline{10-11}\cline{13-15}\cline{17-19}
&  & PW$^{a}$ & LYP$^{b}$ &  & WF$^{c}$ & SPCR$^{d}$ & ECI$^{e}$ &  & PW$%
^{a} $ & LYP$^{b}$ &  & WF$^{c}$ & ECI$^{e}$ & RRV$^{f}$ &  & ZBS$^{g}$ & RPR%
$^{h} $ & RBB$^{i}$ \\ \hline
\multicolumn{1}{r}{$1s2s2p$ $^{4}P$} &  & \multicolumn{1}{l}{$5.3811$} & 
\multicolumn{1}{l}{$5.4117$} & \multicolumn{1}{l}{} & \multicolumn{1}{l}{
5.4114} & \multicolumn{1}{l}{5.3678} & \multicolumn{1}{l}{5.3679} & 
\multicolumn{1}{l}{} & \multicolumn{1}{l}{$57.2151$} & \multicolumn{1}{l}{$%
56.4777$} & \multicolumn{1}{l}{} & \multicolumn{1}{l}{56.728} & 
\multicolumn{1}{l}{57.413} & \multicolumn{1}{l}{57.47} & \multicolumn{1}{l}{}
& \multicolumn{1}{l}{57.385} & \multicolumn{1}{l}{57.41} & 
\multicolumn{1}{l}{57.432} \\ 
\multicolumn{1}{r}{$^{2}P$} &  & \multicolumn{1}{l}{$5.3177^{\ast }$} & 
\multicolumn{1}{l}{$5.2854^{\ast }$} & \multicolumn{1}{l}{} & 
\multicolumn{1}{l}{5.2912} & \multicolumn{1}{l}{5.3133$^{\star }$} & 
\multicolumn{1}{l}{} & \multicolumn{1}{l}{} & \multicolumn{1}{l}{$%
58.9403^{\ast }$} & \multicolumn{1}{l}{$59.9145^{\ast }$} & 
\multicolumn{1}{l}{} & \multicolumn{1}{l}{59.999} & \multicolumn{1}{l}{} & 
\multicolumn{1}{l}{} & \multicolumn{1}{l}{} & \multicolumn{1}{l}{58.912$%
^{\star }$} & \multicolumn{1}{l}{58.912$^{\star }$} & \multicolumn{1}{l}{
58.912$^{\star }$} \\ 
\multicolumn{1}{r}{} &  & \multicolumn{1}{l}{$5.2407^{\dagger }$} & 
\multicolumn{1}{l}{$5.2223^{\dagger }$} & \multicolumn{1}{l}{} & 
\multicolumn{1}{l}{} & \multicolumn{1}{l}{5.2585$^{\dagger }$} & 
\multicolumn{1}{l}{} & \multicolumn{1}{l}{} & \multicolumn{1}{l}{$%
61.0356^{\dagger }$} & \multicolumn{1}{l}{$61.6316^{\dagger }$} & 
\multicolumn{1}{l}{} & \multicolumn{1}{l}{} & \multicolumn{1}{l}{} & 
\multicolumn{1}{l}{} & \multicolumn{1}{l}{} & \multicolumn{1}{l}{60.397$%
^{\dagger }$} & \multicolumn{1}{l}{60.405$^{\dagger }$} & \multicolumn{1}{l}{
60.402$^{\dagger }$} \\ 
\multicolumn{1}{r}{$1s2s3p$ $^{4}P$} &  & \multicolumn{1}{l}{$5.1994$} & 
\multicolumn{1}{l}{$5.2212$} & \multicolumn{1}{l}{} & \multicolumn{1}{l}{
5.2205} & \multicolumn{1}{l}{} & \multicolumn{1}{l}{5.1873} & 
\multicolumn{1}{l}{} & \multicolumn{1}{l}{$62.1595$} & \multicolumn{1}{l}{$%
61.6610$} & \multicolumn{1}{l}{} & \multicolumn{1}{l}{61.923} & 
\multicolumn{1}{l}{61.262} & \multicolumn{1}{l}{61.33} & \multicolumn{1}{l}{}
& \multicolumn{1}{l}{} & \multicolumn{1}{l}{} & \multicolumn{1}{l}{} \\ 
\multicolumn{1}{r}{$^{2}P$} & \multicolumn{1}{l}{} & \multicolumn{1}{l}{$%
5.1589$} & \multicolumn{1}{l}{$5.1308$} & \multicolumn{1}{l}{} & 
\multicolumn{1}{l}{5.1767} & \multicolumn{1}{l}{5.1841} & \multicolumn{1}{l}{
} & \multicolumn{1}{l}{} & \multicolumn{1}{l}{$63.2615$} & 
\multicolumn{1}{l}{$64.1220$} & \multicolumn{1}{l}{} & \multicolumn{1}{l}{
63.115} & \multicolumn{1}{l}{} & \multicolumn{1}{l}{} & \multicolumn{1}{l}{}
& \multicolumn{1}{l}{62.425} & \multicolumn{1}{l}{62.421} & 
\multicolumn{1}{l}{62.462} \\ 
\multicolumn{1}{r}{$1s2s4p$ $^{4}P$} &  & \multicolumn{1}{l}{$5.1623$} & 
\multicolumn{1}{l}{$5.1835$} & \multicolumn{1}{l}{} & \multicolumn{1}{l}{} & 
\multicolumn{1}{l}{} & \multicolumn{1}{l}{} & \multicolumn{1}{l}{} & 
\multicolumn{1}{l}{$63.1690$} & \multicolumn{1}{l}{$62.6874$} & 
\multicolumn{1}{l}{} & \multicolumn{1}{l}{} & \multicolumn{1}{l}{} & 
\multicolumn{1}{l}{} & \multicolumn{1}{l}{} & \multicolumn{1}{l}{} & 
\multicolumn{1}{l}{} & \multicolumn{1}{l}{} \\ 
\multicolumn{1}{r}{$^{2}P$} &  & \multicolumn{1}{l}{$5.1232$} & 
\multicolumn{1}{l}{$5.0973$} & \multicolumn{1}{l}{} & \multicolumn{1}{l}{} & 
\multicolumn{1}{l}{} & \multicolumn{1}{l}{} & \multicolumn{1}{l}{} & 
\multicolumn{1}{l}{$64.2330$} & \multicolumn{1}{l}{$65.0330$} & 
\multicolumn{1}{l}{} & \multicolumn{1}{l}{} & \multicolumn{1}{l}{} & 
\multicolumn{1}{l}{} & \multicolumn{1}{l}{} & \multicolumn{1}{l}{} & 
\multicolumn{1}{l}{} & \multicolumn{1}{l}{} \\ 
\multicolumn{1}{r}{$1s2s5p$ $^{4}P$} &  & \multicolumn{1}{l}{$5.1470$} & 
\multicolumn{1}{l}{$5.1676$} & \multicolumn{1}{l}{} & \multicolumn{1}{l}{} & 
\multicolumn{1}{l}{} & \multicolumn{1}{l}{} & \multicolumn{1}{l}{} & 
\multicolumn{1}{l}{$63.5853$} & \multicolumn{1}{l}{$63.1200$} & 
\multicolumn{1}{l}{} & \multicolumn{1}{l}{} & \multicolumn{1}{l}{} & 
\multicolumn{1}{l}{} & \multicolumn{1}{l}{} & \multicolumn{1}{l}{} & 
\multicolumn{1}{l}{} & \multicolumn{1}{l}{} \\ 
\multicolumn{1}{r}{$^{2}P$} &  & \multicolumn{1}{l}{$5.1081$} & 
\multicolumn{1}{l}{$5.0832$} & \multicolumn{1}{l}{} & \multicolumn{1}{l}{} & 
\multicolumn{1}{l}{} & \multicolumn{1}{l}{} & \multicolumn{1}{l}{} & 
\multicolumn{1}{l}{$64.6439$} & \multicolumn{1}{l}{$65.4167$} & 
\multicolumn{1}{l}{} & \multicolumn{1}{l}{} & \multicolumn{1}{l}{} & 
\multicolumn{1}{l}{} & \multicolumn{1}{l}{} & \multicolumn{1}{l}{} & 
\multicolumn{1}{l}{} & \multicolumn{1}{l}{} \\ 
\multicolumn{1}{r}{$1s2s6p$ $^{4}P$} &  & \multicolumn{1}{l}{$5.1391$} & 
\multicolumn{1}{l}{$5.1594$} & \multicolumn{1}{l}{} & \multicolumn{1}{l}{} & 
\multicolumn{1}{l}{} & \multicolumn{1}{l}{} & \multicolumn{1}{l}{} & 
\multicolumn{1}{l}{$63.8003$} & \multicolumn{1}{l}{$63.3432$} & 
\multicolumn{1}{l}{} & \multicolumn{1}{l}{} & \multicolumn{1}{l}{} & 
\multicolumn{1}{l}{} & \multicolumn{1}{l}{} & \multicolumn{1}{l}{} & 
\multicolumn{1}{l}{} & \multicolumn{1}{l}{} \\ 
\multicolumn{1}{r}{$^{2}P$} &  & \multicolumn{1}{l}{$5.1003$} & 
\multicolumn{1}{l}{$5.0759$} & \multicolumn{1}{l}{} & \multicolumn{1}{l}{} & 
\multicolumn{1}{l}{} & \multicolumn{1}{l}{} & \multicolumn{1}{l}{} & 
\multicolumn{1}{l}{$64.8561$} & \multicolumn{1}{l}{$65.6153$} & 
\multicolumn{1}{l}{} & \multicolumn{1}{l}{} & \multicolumn{1}{l}{} & 
\multicolumn{1}{l}{} & \multicolumn{1}{l}{} & \multicolumn{1}{l}{} & 
\multicolumn{1}{l}{} & \multicolumn{1}{l}{} \\ 
\multicolumn{1}{r}{$1s2s7p$ $^{4}P$} &  & \multicolumn{1}{l}{$5.1345$} & 
\multicolumn{1}{l}{$5.1547$} & \multicolumn{1}{l}{} & \multicolumn{1}{l}{} & 
\multicolumn{1}{l}{} & \multicolumn{1}{l}{} & \multicolumn{1}{l}{} & 
\multicolumn{1}{l}{$63.9255$} & \multicolumn{1}{l}{$63.4711$} & 
\multicolumn{1}{l}{} & \multicolumn{1}{l}{} & \multicolumn{1}{l}{} & 
\multicolumn{1}{l}{} & \multicolumn{1}{l}{} & \multicolumn{1}{l}{} & 
\multicolumn{1}{l}{} & \multicolumn{1}{l}{} \\ 
\multicolumn{1}{r}{$^{2}P$} &  & \multicolumn{1}{l}{$5.0956$} & 
\multicolumn{1}{l}{$5.0717$} & \multicolumn{1}{l}{} & \multicolumn{1}{l}{} & 
\multicolumn{1}{l}{} & \multicolumn{1}{l}{} & \multicolumn{1}{l}{} & 
\multicolumn{1}{l}{$64.9840$} & \multicolumn{1}{l}{$65.7296$} & 
\multicolumn{1}{l}{} & \multicolumn{1}{l}{} & \multicolumn{1}{l}{} & 
\multicolumn{1}{l}{} & \multicolumn{1}{l}{} & \multicolumn{1}{l}{} & 
\multicolumn{1}{l}{} & \multicolumn{1}{l}{} \\ 
\multicolumn{1}{r}{$1s2s8p$ $^{4}P$} &  & \multicolumn{1}{l}{$5.1315$} & 
\multicolumn{1}{l}{$5.1516$} & \multicolumn{1}{l}{} & \multicolumn{1}{l}{} & 
\multicolumn{1}{l}{} & \multicolumn{1}{l}{} & \multicolumn{1}{l}{} & 
\multicolumn{1}{l}{$64.0071$} & \multicolumn{1}{l}{$63.5554$} & 
\multicolumn{1}{l}{} & \multicolumn{1}{l}{} & \multicolumn{1}{l}{} & 
\multicolumn{1}{l}{} & \multicolumn{1}{l}{} & \multicolumn{1}{l}{} & 
\multicolumn{1}{l}{} & \multicolumn{1}{l}{} \\ 
\multicolumn{1}{r}{$^{2}P$} &  & \multicolumn{1}{l}{$5.0928$} & 
\multicolumn{1}{l}{$5.0690$} & \multicolumn{1}{l}{} & \multicolumn{1}{l}{} & 
\multicolumn{1}{l}{} & \multicolumn{1}{l}{} & \multicolumn{1}{l}{} & 
\multicolumn{1}{l}{$65.0602$} & \multicolumn{1}{l}{$65.8031$} & 
\multicolumn{1}{l}{} & \multicolumn{1}{l}{} & \multicolumn{1}{l}{} & 
\multicolumn{1}{l}{} & \multicolumn{1}{l}{} & \multicolumn{1}{l}{} & 
\multicolumn{1}{l}{} & \multicolumn{1}{l}{} \\ \hline\hline
\end{tabular}

$^{a}$PW results, $^{b}$LYP results, $^{c}$\cite{Roy97b}, $^{d}$\cite%
{Chen94,Davis89}, $^{e}$\cite{Bunge78,Bunge78-1,Bunge81}, $^{f}$\cite%
{Holoien67}, $^{g}$\cite{Ziem75}, $^{h}$\cite{Rassi77}, and $^{i}$\cite%
{Rodbro79}. Here, $^{\star }$for $1s(2s2p$ $^{3}P)$ $^{2}P$ and $^{\dagger }$%
for $1s(2s2p$ $^{1}P)$ $^{2}P$.%
\end{table*}%
\endgroup

\begingroup\squeezetable

\begin{table*}[htbp] \centering%
\caption{Total energies ($E$) and excitation energies ($\Delta E$) of inner-shell doubly 
excited states $1s2p^2$ $^2S$, $^{2,4}P$, and $^2D$ of Li.\label{T-3}}%
\begin{tabular}{ccccccccccccccccccc}
\hline\hline
& \ \  &  & $-E$ &  &  & (a.u.) &  & \ \  &  &  &  & $\Delta E$ &  &  &  & 
(eV) &  &  \\ \cline{3-8}\cline{10-19}
States &  & Present & work &  & \multicolumn{1}{l}{Other} &  & theory &  & 
Present & work &  & Other &  & theory &  &  & Exp. &  \\ 
\cline{3-4}\cline{6-8}\cline{10-11}\cline{13-15}\cline{17-19}
&  & PW$^{a}$ & LYP$^{b}$ &  & WF$^{c}$ & SPCR$^{d}$ & ECI$^{e}$ &  & PW$%
^{a} $ & LYP$^{b}$ &  & WF$^{c}$ & ECI$^{e}$ & RRV$^{f}$ &  & ZBS$^{g}$ & RPR%
$^{h} $ & RBB$^{i}$ \\ \hline
\multicolumn{1}{r}{$1s2p^{2}$ $^{4}P$} &  & $5.2526$ & $5.2862$ &  & 5.2860
& 5.2453 & 5.2453 &  & $60.7118$ & $59.8927$ &  & 60.140 & 60.750 & 60.74 & 
&  & 60.75 & 60.802 \\ 
\multicolumn{1}{r}{$^{2}D$} &  & $5.2254$ & $5.2232$ &  & 5.2356 & 5.2342 & 
&  & $61.4520$ & $61.6071$ &  & 61.512 &  &  &  & 61.065 & 61.065 & 61.062
\\ 
\multicolumn{1}{r}{$^{2}P$} &  & $5.2382$ & $5.1930$ &  & 5.2323 & 5.2137 & 
5.2137 &  & $61.1036$ & $62.4289$ &  & 61.602 &  &  &  &  &  &  \\ 
\multicolumn{1}{r}{$^{2}S$} & \multicolumn{1}{l}{} & $5.1761$ & $5.1739$ & 
& 5.1780 &  &  &  & $62.7935$ & $62.9486$ &  & 63.079 &  &  &  &  &  & 63.492
\\ \hline\hline
\end{tabular}

$^{a}$PW results, $^{b}$LYP results, $^{c}$\cite{Roy97b}, $^{d}$\cite%
{Davis88,Chen94}, $^{e}$\cite{Bunge78,Bunge78-1,Bunge79}, $^{f}$\cite%
{Holoien67},$^{g}$\cite{Ziem75}, $^{h}$\cite{Rassi77}, and $^{i}$\cite%
{Rodbro79}.%
\end{table*}%
\endgroup

\subsection{Inner-shell excitation of B}

The second open-shell atom for which the total energies of inner-shell
excited states are computed is B. In TABLE \ref{T-4} we present Auger
transition energies from the inner-shell excited states $1s2s^{2}2p^{2}$ $%
^{2}S$, $^{2,4}P$, and $^{2}D$ of B to the singly excited states $1s^{2}2s2p$
$^{1,3}P$ and the doubly excited states $1s^{2}2p^{2}$ $^{1}S$, $^{3}P$, and 
$^{1}D$ \ of B$^{+}$. For comparison we also give in this table the
available experimental results \cite{Rodbro79} as well as theoretical
results from WF \cite{Roy97b}, perturbation Z-expansion theory (PZE) \cite%
{Safronova69,Ivanova75}, and perturbation Z-expansion theory with
relativistic effects (RPZE) \cite{Rodbro79}. Except for the transitions $%
1s2s^{2}2p^{2}$ $^{2}D\rightarrow 1s^{2}2p^{2\text{ \ }1}S$ and $%
1s2s^{2}2p^{2}$ $^{2}S\rightarrow 1s^{2}2p^{2\text{ \ }1}S$, the maximum
deviations of our PW and LYP results to the experimental results are 0.634\%
and 0.241\%, while the maximum discrepancies of the WF, PZE, and RPZE
results to the experimental results are 0.783\%, 1.211\%, and 0.4091\%,
respectively. This indicates that our results are much better than the PZE
results, a little bit better than the WF results, and much closer to the
RPZE results. The maximum deviations of the PW, LYP, PZE, and RPZE results
to the experimental results are 1.821\%, 1.033\%, 3.257\%, and 0.07\% for
the transition $1s2s^{2}2p^{2}$ $^{2}D\rightarrow 1s^{2}2p^{2\text{ \ }1}S$
and 0.437\%, 1.232\%, 1.392\%, and 0.04\% for the transition $1s2s^{2}2p^{2}$
$^{2}S\rightarrow 1s^{2}2p^{2\text{ \ }1}S$, respectively. This indicates
again that our results are better than the PZE results. In addition,
according to our result the spectrum line with 170.7 eV previously signed to
the transition $1s2s^{2}2p^{2}$ $^{2}D\rightarrow 1s^{2}2p^{2\text{ \ }1}S$
may likely belong to the transition $1s2s^{2}2p^{2}$ $^{2}D\rightarrow
1s^{2}2p^{2\text{ \ }3}P$ .

\begingroup\squeezetable

\begin{table*}[htbp] \centering%
\caption{Auger transition energies ($\Delta E$) from inner-shell excited states 
$1s2s^22p^2$ $^2S$, $^{2,4}P$, and $^2D$ of B to singly excited states 
$1s^22s2p$ $^{1,3}P$ and doubly excited states 
$1s^22p^2$ $^1S$,  $^3P$, and  $^1D$ of B$^+$.\label{T-4}}%
\begin{tabular}{ccccccccccccccc}
\hline\hline
& \ \ \  &  & \ \ \ \ \ \  &  &  & $\Delta E$ & \ \  &  &  &  &  & (eV) & \
\  &  \\ \cline{5-15}
Initial states &  & Final states &  & Present &  & work &  & Other &  &  & 
& theory &  & Experiment \\ \cline{5-7}\cline{9-13}\cline{15-15}
&  &  &  & PW$^{a}$ &  & LYP$^{b}$ &  & WF$^{c}$ &  & PZE$^{d}$ &  & RPZE$%
^{e}$ &  & RBB$^{f}$ \\ \hline
\multicolumn{1}{r}{$1s2s^{2}2p^{2}$ $^{4}P$} & \multicolumn{1}{r}{} & 
\multicolumn{1}{l}{$1s^{2}2p^{2}$ $^{3}P$} &  & 168.3500 &  & 169.1008 &  & 
&  & $167.4101$ &  & 169.10 &  & 169.2 \\ 
\multicolumn{1}{r}{} & \multicolumn{1}{r}{} & \multicolumn{1}{l}{$%
1s^{2}2p^{2}$ $^{1}D$} &  & 167.3105 &  & 168.5612 &  &  &  & $166.7952$ & 
&  &  &  \\ 
\multicolumn{1}{r}{} & \multicolumn{1}{r}{} & \multicolumn{1}{l}{$%
1s^{2}2p^{2}$ $^{1}S$} &  & 165.5064 &  & 167.2558 &  &  &  & $163.2454$ & 
&  &  &  \\ 
\multicolumn{1}{r}{$^{2}D$} & \multicolumn{1}{r}{} & \multicolumn{1}{l}{$%
1s^{2}2p^{2}$ $^{3}P$} &  & 170.4344 &  & 170.7805 &  &  &  & $169.3045$ & 
&  &  &  \\ 
\multicolumn{1}{r}{} & \multicolumn{1}{r}{} & \multicolumn{1}{l}{$%
1s^{2}2p^{2}$ $^{1}D$} &  & 169.3949 &  & 170.2409 &  &  &  & $168.6895$ & 
&  &  &  \\ 
\multicolumn{1}{r}{} & \multicolumn{1}{r}{} & \multicolumn{1}{l}{$%
1s^{2}2p^{2}$ $^{1}S$} &  & 167.5908 &  & 168.9356 &  &  &  & $165.1397$ & 
& 170.58 &  & 170.7 \\ 
\multicolumn{1}{r}{$^{2}P$} & \multicolumn{1}{r}{} & \multicolumn{1}{l}{$%
1s^{2}2p^{2}$ $^{3}P$} &  & 170.4181 &  & 170.2883 &  & 171.5858 &  & $%
169.0290$ &  & 170.73 &  & 170.7 \\ 
\multicolumn{1}{r}{} & \multicolumn{1}{r}{} & \multicolumn{1}{l}{$%
1s^{2}2p^{2}$ $^{1}D$} &  & 169.3786 &  & 169.7487 &  &  &  & $168.4140$ & 
&  &  &  \\ 
\multicolumn{1}{r}{} & \multicolumn{1}{r}{} & \multicolumn{1}{l}{$%
1s^{2}2p^{2}$ $^{1}S$} &  & 167.5745 &  & 168.4433 &  &  &  & $164.8643$ & 
&  &  &  \\ 
\multicolumn{1}{r}{$^{2}S$} & \multicolumn{1}{r}{} & \multicolumn{1}{l}{$%
1s^{2}2p^{2}$ $^{3}P$} &  & 172.7828 &  & 173.1289 &  &  &  & $171.0097$ & 
&  &  &  \\ 
\multicolumn{1}{r}{} & \multicolumn{1}{r}{} & \multicolumn{1}{l}{$%
1s^{2}2p^{2}$ $^{1}D$} &  & 171.7433 &  & 172.5893 &  &  &  & $170.3947$ & 
&  &  &  \\ 
\multicolumn{1}{r}{} & \multicolumn{1}{r}{} & \multicolumn{1}{l}{$%
1s^{2}2p^{2}$ $^{1}S$} &  & 169.9392 &  & 171.2840 &  &  &  & $166.8450$ & 
& 169.13 &  & 169.2 \\ 
\multicolumn{1}{r}{} & \multicolumn{1}{r}{} & \multicolumn{1}{l}{} &  &  & 
&  &  &  &  &  &  &  &  &  \\ 
\multicolumn{1}{r}{$1s2s^{2}2p^{2}$ $^{4}P$} & \multicolumn{1}{r}{} & 
\multicolumn{1}{l}{$1s^{2}2s2p$ $^{3}P$} &  & 176.1516 &  & 176.8376 &  &  & 
& $175.7192$ &  &  &  &  \\ 
\multicolumn{1}{r}{} & \multicolumn{1}{r}{} & \multicolumn{1}{l}{$1s^{2}2s2p$
$^{1}P$} &  & 171.4249 &  & 172.9961 &  &  &  & $170.8674$ &  &  &  &  \\ 
\multicolumn{1}{r}{$^{2}D$} & \multicolumn{1}{r}{} & \multicolumn{1}{l}{$%
1s^{2}2s2p$ $^{3}P$} &  & 178.2360 &  & 178.5173 &  &  &  & $177.6135$ &  & 
&  &  \\ 
\multicolumn{1}{r}{} & \multicolumn{1}{r}{} & \multicolumn{1}{l}{$1s^{2}2s2p$
$^{1}P$} &  & 173.5093 &  & 174.6759 &  & 173.2321 &  & $172.7617$ &  & 
174.17 &  & 174.6 \\ 
\multicolumn{1}{r}{$^{2}P$} & \multicolumn{1}{r}{} & \multicolumn{1}{l}{$%
1s^{2}2s2p$ $^{3}P$} &  & 178.2197 &  & 178.0251 &  &  &  & $177.3381$ &  & 
&  &  \\ 
\multicolumn{1}{r}{} & \multicolumn{1}{r}{} & \multicolumn{1}{l}{$1s^{2}2s2p$
$^{1}P$} &  & 173.4930 &  & 174.1836 &  & 173.4062 &  & $172.4862$ &  & 
173.90 &  & 174.6 \\ 
\multicolumn{1}{r}{$^{2}S$} & \multicolumn{1}{r}{} & \multicolumn{1}{l}{$%
1s^{2}2s2p$ $^{3}P$} &  & 180.5843 &  & 180.8657 &  &  &  & $179.3188$ &  & 
&  &  \\ 
\multicolumn{1}{r}{} & \multicolumn{1}{r}{} & \multicolumn{1}{l}{$1s^{2}2s2p$
$^{1}P$} &  & 175.8577 &  & 177.0242 &  &  &  & $174.4669$ &  &  &  &  \\ 
\hline\hline
\end{tabular}

$^{a}$PW results, $^{b}$LYP results, $^{c}$\cite{Roy97b}, $^{d}$\cite%
{Safronova69,Ivanova75}, $^{e}$\cite{Rodbro79}, and $^{f}$Ref. \cite%
{Rodbro79}. 
\end{table*}%
\endgroup

\subsection{Inner-shell excitation of positive ions Ne$^{+}$, Ne$^{2+}$, and
Ne$^{+}$}

To explore the feasibility of the approach to inner-shell excitation of
atomic ions, we also apply the procedure to inner-shell excited-state
calculations of positive ions Ne$^{+}$, Ne$^{2+}$, and Ne$^{3+}$. In TABLE %
\ref{T-5} we present the excitation energies of optically allowed
transitions involved in inner-shell excited states of these ions along with
the theoretical results of multiconfiguration Dirac-Fock (MCDF) method \cite%
{Yamaoka2001}. Note that the MCDF results here are the weighted-averaged
values of those given in \cite{Yamaoka2001}. The maximum relative deviations
of our PW and LYP results to the MCDF results are 0.171\% and 0.071\% for Ne$%
^{+}$, 0.162\% and 0.059\% for Ne$^{2+}$, and 0.159\% and 0.081\% for Ne$%
^{3+}$, respectively. Thus the agreement of our results with the MCDF
results are quite satisfactory, demonstrating that the SLHF together with PW
and LYP correlation potentials is accurate for the calculation of
inner-shell excited states of Ne positive ions. In addition, the LYP results
are more accurate than the PW results for these atomic systems with $Z=10$ 
\cite{zhou2005,zhou06-inner-close-shell}.

\begingroup\squeezetable

\begin{table}[htbp] \centering%
\caption{Excitation energies ($\Delta E$) for optically allowed transitions 
of inner-shell excited states of Ne$^{k+}$ ($k=1 \sim 3$).\label{T-5}}%
\begin{tabular}{ccccccccccc}
\hline\hline
&  &  &  &  &  & $\Delta E$ &  &  &  & (eV) \\ \cline{7-11}
Ions &  & Initial State &  & Final State &  & Present &  & work &  & Others
\\ \cline{7-9}\cline{11-11}
&  &  &  &  &  & PW$^{a}$ &  & LYP$^{b}$ &  & MCDF$^{c}$ \\ \hline
Ne$^{+}$ &  & \multicolumn{1}{r}{$1s^{2}2s^{2}2p^{5}$ $^{2}P$} & 
\multicolumn{1}{r}{} & \multicolumn{1}{r}{$1s2s^{2}2p^{6}$ $^{2}S$} &  & 
848.0278 &  & 848.8786 &  & 849.48 \\ 
&  & \multicolumn{1}{r}{} & \multicolumn{1}{r}{} & \multicolumn{1}{r}{} &  & 
&  &  &  &  \\ 
Ne$^{2+}$ &  & \multicolumn{1}{r}{$1s^{2}2s^{2}2p^{4\text{ \ }3}P$} & 
\multicolumn{1}{r}{} & \multicolumn{1}{r}{$1s2s^{2}2p^{5\text{ \ }3}P$} &  & 
853.7724 &  & 854.6535 &  & 855.16 \\ 
&  & \multicolumn{1}{r}{$^{1}D$} & \multicolumn{1}{r}{} & \multicolumn{1}{r}{%
$^{1}P$} &  & 855.0747 &  & 856.3277 &  & 856.14 \\ 
&  & \multicolumn{1}{r}{$^{1}S$} & \multicolumn{1}{r}{} & \multicolumn{1}{r}{%
$^{1}P$} &  & 850.1188 &  & 851.3728 &  & 851.19 \\ 
&  & \multicolumn{1}{r}{} & \multicolumn{1}{r}{} & \multicolumn{1}{r}{} &  & 
&  &  &  &  \\ 
Ne$^{3+}$ &  & \multicolumn{1}{r}{$1s^{2}2s^{2}2p^{3\text{ \ }4}S$} & 
\multicolumn{1}{r}{} & \multicolumn{1}{r}{$1s2s^{2}2p^{4\text{ \ }4}P$} &  & 
861.0411 &  & 861.8325 &  & 862.41 \\ 
&  & \multicolumn{1}{r}{$^{2}D$} & \multicolumn{1}{r}{} & \multicolumn{1}{r}{%
$^{2}D$} &  & 862.2075 &  & 863.0207 &  & 863.33 \\ 
&  & \multicolumn{1}{r}{} & \multicolumn{1}{r}{} & \multicolumn{1}{r}{$^{2}P$%
} &  & 863.1715 &  & 863.6460 &  & 864.20 \\ 
&  & \multicolumn{1}{r}{$^{2}P$} & \multicolumn{1}{r}{} & \multicolumn{1}{r}{%
$^{2}D$} &  & 858.5203 &  & 859.3295 &  & 859.78 \\ 
&  & \multicolumn{1}{r}{} & \multicolumn{1}{r}{} & \multicolumn{1}{r}{$^{2}P$%
} &  & 859.4843 &  & 859.9549 &  & 860.65 \\ 
&  & \multicolumn{1}{r}{} & \multicolumn{1}{r}{} & \multicolumn{1}{r}{$^{2}S$%
} &  & 864.5179 &  & 865.3302 &  & 865.62 \\ \hline\hline
\end{tabular}

$^{a}$PW results, $^{b}$LYP results, and $^{c}$obtained with data in \cite%
{Yamaoka2001}.%
\end{table}%
\endgroup

\subsection{Inner-shell excitation of Na}

Finally we compute the total energies and excitation energies for
inner-shell excited states $1s2s^{2}2p^{6}3s(^{1}S)np$ $^{2}P$ $\left(
n=3\sim 8\right) $ and $1s2s^{2}2p^{6}3s(^{3}S)np$ $^{2,4}P$ $\left( n=3\sim
8\right) $ of Na. The results are shown in TABLE \ref{T-6} together with the
theoretical results from relativistic configuration interaction method (RCI) 
\cite{Yang93} and multiconfiguration Hartree-Fock method (MCHF) \cite%
{Yavna86} as well as the experimental results (Exp.) \cite{Tuilier82} for
comparison.

For the excitation energies of excited states $1s2s^{2}2p^{6}3s(^{1}S)np$ $%
^{2}P$, the relative deviations of our PW and LYP results to the
experimental results are not more than 0.18\% and 0.07\%, respectively. This
indicates that both the PW and LYP results are in good agreement with the
experimental results and the LYP results are better than the PW results. In
addition, the maximum relative discrepancies of our RCI and MCHF results to
the experimental results are 0.06\% and 0.04\%, respectively. Thus our LYP
results are very close to the RCI results and MCHF results. Furthermore,
according to our results the identification of the photoionization spectra $%
1s2s^{2}2p^{6}3s(^{1}S)3p$ $^{2}P$ and $1s2s^{2}2p^{6}3s(^{3}S)3p$ $^{2}P$
in the experiment \cite{Tuilier82} should be exchanged. This result agrees
well with the theoretical results of \cite{Yavna86} and \cite{Yang93}.

For the excitation energy of excited state $1s2s^{2}2p^{6}3s(^{3}S)3p$ $%
^{2}P $, the maximum relative deviations of our PW and LYP results to the
experimental results are 0.22\% and 0.16\%, respectively, which are a little
bit larger than those of RCI and MCHF results to the experimental results.
For the excitation energies of excited states $1s2s^{2}2p^{6}3s(^{3}S)np$ $%
^{2}P$ with $n=4$ and $5$, the maximum relative deviations of our PW and LYP
results are 0.12\% and 0.04\% to the RCI results, and 0.18\% and 0.10\% to
the MCHF results, respectively. This demonstrates that our LYP results are
very close to the RCI results for these states.

For the excitation energies of excited states $1s2s^{2}2p^{6}3s(^{3}S)np$ $%
^{4}P$, the relative deviations of our PW and LYP results to the RCI results
are less than 0.13\% and 0.03\%, respectively. This illustrates again that
our LYP results are in very agreement with the RCI results for these states.
For all the excited states the LYP results are better than the PW results.
Thus the LYP energy functional is more accurate than the PW energy
functional in the excitation energy calculation of inner-shell excited
states of atomic systems with large $Z$.

\begingroup\squeezetable

\begin{table*}[htbp] \centering%
\caption{Total energies ($E$) and excitation energies ($\Delta E$) of inner-shell 
excited states $1s2s^22p^63s(^1S)np$ $^2P$ and 
$1s2s^22p^63s(^3S)np$ $^{2,4}P$ ($n=3\thicksim 8$) of Na. 
The ground state energies obtained from calculations with PW and LYP correlation 
potentials and energy functionals are $-162.2265$ (a.u.) and $-162.2687$ (a.u.), 
respectively.\label{T-6}}%
\begin{tabular}{ccccccccccccccc}
\hline\hline
& \ \ \ \ \ \ \  & $-E$ & \  & (a.u.) & \ \ \ \ \ \ \  &  & \  & $\Delta E$
& \ \ \ \  &  &  & (eV) & \ \ \ \  &  \\ \cline{3-5}\cline{7-15}
States &  & Present &  & work &  & Present &  & work &  & Other &  & theory
&  & Exp. \\ \cline{3-5}\cline{7-9}\cline{11-13}\cline{15-15}
&  & PW$^{a}$ &  & LYP$^{b}$ &  & PW$^{a}$ &  & LYP$^{b}$ &  & RCI$^{c}$ & 
& MCHF$^{d}$ &  & TLE$^{e}$ \\ \hline
\multicolumn{1}{r}{$1s2s^{2}2p^{6}3s(^{1}S)3p$ $^{2}P$} &  & 
\multicolumn{1}{l}{$122.7947$} &  & \multicolumn{1}{l}{$122.7552$} &  & 
\multicolumn{1}{l}{$1073.0024$} &  & \multicolumn{1}{l}{$1074.1765$} & 
\multicolumn{1}{l}{} & 1074.28 &  & 1074.50 & \multicolumn{1}{l}{} & 1074.95
\\ 
\multicolumn{1}{r}{$4p$ $^{2}P$} &  & \multicolumn{1}{l}{$122.6680$} &  & 
\multicolumn{1}{l}{$122.6708$} &  & \multicolumn{1}{l}{$1076.4501$} &  & 
\multicolumn{1}{l}{$1077.5222$} & \multicolumn{1}{l}{} & 1077.52 &  & 1077.96
& \multicolumn{1}{l}{} & 1078.17 \\ 
\multicolumn{1}{r}{$5p$ $^{2}P$} &  & \multicolumn{1}{l}{$122.6378$} &  & 
\multicolumn{1}{l}{$122.6424$} &  & \multicolumn{1}{l}{$1077.2719$} &  & 
\multicolumn{1}{l}{$1078.2950$} & \multicolumn{1}{l}{} & 1078.27 &  & 1078.60
& \multicolumn{1}{l}{} & 1078.9 \\ 
\multicolumn{1}{r}{$6p$ $^{2}P$} &  & \multicolumn{1}{l}{$122.6246$} &  & 
\multicolumn{1}{l}{$122.6301$} &  & \multicolumn{1}{l}{$1077.6311$} &  & 
\multicolumn{1}{l}{$1078.6297$} & \multicolumn{1}{l}{} &  &  &  & 
\multicolumn{1}{l}{} &  \\ 
\multicolumn{1}{r}{$7p$ $^{2}P$} &  & \multicolumn{1}{l}{$122.6176$} &  & 
\multicolumn{1}{l}{$122.6234$} &  & \multicolumn{1}{l}{$1077.8215$} &  & 
\multicolumn{1}{l}{$1078.8120$} & \multicolumn{1}{l}{} &  &  &  & 
\multicolumn{1}{l}{} &  \\ 
\multicolumn{1}{r}{$8p$ $^{2}P$} &  & \multicolumn{1}{l}{$122.6134$} &  & 
\multicolumn{1}{l}{$122.6195$} &  & \multicolumn{1}{l}{$1077.9358$} &  & 
\multicolumn{1}{l}{$1078.9182$} & \multicolumn{1}{l}{} &  &  &  & 
\multicolumn{1}{l}{} &  \\ 
\multicolumn{1}{r}{} &  & \multicolumn{1}{l}{} &  & \multicolumn{1}{l}{} & 
& \multicolumn{1}{l}{} &  & \multicolumn{1}{l}{} & \multicolumn{1}{l}{} &  & 
&  & \multicolumn{1}{l}{} &  \\ 
\multicolumn{1}{r}{$1s2s^{2}2p^{6}3s(^{3}S)3p$ $^{2}P$} &  & 
\multicolumn{1}{l}{$122.7540$} &  & \multicolumn{1}{l}{$122.7714$} &  & 
\multicolumn{1}{l}{$1074.1099$} &  & \multicolumn{1}{l}{$1074.7847$} & 
\multicolumn{1}{l}{} & 1075.97 &  & 1076.04 & \multicolumn{1}{l}{} & 1076.47
\\ 
\multicolumn{1}{r}{$4p$ $^{2}P$} &  & \multicolumn{1}{l}{$122.6858$} &  & 
\multicolumn{1}{l}{$122.6976$} &  & \multicolumn{1}{l}{$1075.9657$} &  & 
\multicolumn{1}{l}{$1076.7929$} & \multicolumn{1}{l}{} & 1077.21 &  & 1077.83
& \multicolumn{1}{l}{} &  \\ 
\multicolumn{1}{r}{$5p$ $^{2}P$} &  & \multicolumn{1}{l}{$122.6601$} &  & 
\multicolumn{1}{l}{$122.6705$} &  & \multicolumn{1}{l}{$1076.6651$} &  & 
\multicolumn{1}{l}{$1077.5304$} & \multicolumn{1}{l}{} & 1077.88 &  & 1078.56
& \multicolumn{1}{l}{} &  \\ 
\multicolumn{1}{r}{$6p$ $^{2}P$} &  & \multicolumn{1}{l}{$122.6481$} &  & 
\multicolumn{1}{l}{$122.6582$} &  & \multicolumn{1}{l}{$1076.9916$} &  & 
\multicolumn{1}{l}{$1077.8651$} & \multicolumn{1}{l}{} &  &  &  & 
\multicolumn{1}{l}{} &  \\ 
\multicolumn{1}{r}{$7p$ $^{2}P$} &  & \multicolumn{1}{l}{$122.6414$} &  & 
\multicolumn{1}{l}{$122.6517$} &  & \multicolumn{1}{l}{$1077.1739$} &  & 
\multicolumn{1}{l}{$1078.0420$} & \multicolumn{1}{l}{} &  &  &  & 
\multicolumn{1}{l}{} &  \\ 
\multicolumn{1}{r}{$8p$ $^{2}P$} &  & \multicolumn{1}{l}{$122.6374$} &  & 
\multicolumn{1}{l}{$122.6478$} &  & \multicolumn{1}{l}{$1077.2828$} &  & 
\multicolumn{1}{l}{$1078.1481$} & \multicolumn{1}{l}{} &  &  &  & 
\multicolumn{1}{l}{} &  \\ 
\multicolumn{1}{r}{} &  &  &  &  &  &  &  &  &  &  &  &  &  &  \\ 
\multicolumn{1}{r}{$3p$ $^{4}P$} &  & $122.8152$ &  & $122.8176$ &  & $%
1072.4445$ &  & $1073.5276$ &  & 1073.87 &  &  &  &  \\ 
\multicolumn{1}{r}{$4p$ $^{4}P$} &  & $122.6916$ &  & $122.6983$ &  & $%
1075.8079$ &  & $1076.7739$ &  & 1076.94 &  &  &  &  \\ 
\multicolumn{1}{r}{$5p$ $^{4}P$} &  & $122.6619$ &  & $122.6703$ &  & $%
1076.6161$ &  & $1077.5358$ &  & 1077.71 &  &  &  &  \\ 
\multicolumn{1}{r}{$6p$ $^{4}P$} &  & $122.6487$ &  & $122.6580$ &  & $%
1076.9753$ &  & $1077.8705$ &  &  &  &  &  &  \\ 
\multicolumn{1}{r}{$7p$ $^{4}P$} &  & $122.6418$ &  & $122.6515$ &  & $%
1077.1630$ &  & $1078.0474$ &  &  &  &  &  &  \\ 
\multicolumn{1}{r}{$8p$ $^{4}P$} &  & \multicolumn{1}{l}{$122.6377$} &  & 
\multicolumn{1}{l}{$122.6476$} &  & \multicolumn{1}{l}{$1077.2746$} &  & 
\multicolumn{1}{l}{$1078.1535$} & \multicolumn{1}{l}{} &  &  &  & 
\multicolumn{1}{l}{} &  \\ \hline\hline
\end{tabular}

$^{a}$PW results, $^{b}$LYP results, $^{c}$\cite{Yang93}, $^{d}$\cite%
{Yavna86}, and $^{e}$\cite{Tuilier82}$.$%
\end{table*}%
\endgroup

\section{Conclusions}

In summary, the procedure developed for excited-state calculation based on
the SLHF density functional approach and Slater's diagonal sum rule has been
extended to the treatment of inner-shell excited states of atomic systems.
In this procedure, the electron spin-orbitals in an electronic configuration
are obtained first by solving the KS equation with the exact SLHF exchange
potential. Then a single-Slater-determinant energy of the electronic
configuration is calculated by using these electron spin-orbitals. Finally,
a multiplet energy of an excited state is evaluated from the
single-Slater-determinant energies of the electronic configurations involved
in terms of Slater's diagonal sum rule. The key part of this procedure is
the SLHF exchange potential. We have applied this procedure to the
calculations of total energies and excitation energies of inner-shell
excited states of open-shell atomic systems: Li, B, Ne$^{+}$, Ne$^{2+}$, Ne$%
^{3+}$, and Na. The correlation effect is taken care of by incorporating the
PW and LYP correlation potentials and energy functionals into calculation.
The results from the calculations with LYP and PW energy functionals and
energy functionals are in overall good agreement with each other and also
with the available more sophisticated \textit{ab initio} theoretical results
and experimental data. This demonstrates that the SLHF density-functional
approach can provide a simple and computationally efficient approach for the
accurate calculation of inner-shell excited states of open-shell atomic
systems within DFT.

\bigskip 

\begin{acknowledgments}
This work is partially supported by the Chemical Sciences, Geosciences and
Biosciences Division of the Office of Basic Energy Sciences, Office of
Science, U. S. Department of Energy, and by the National Science Foundation.
\end{acknowledgments}

\bibliographystyle{apsrev}
\bibliography{dft2}

\end{document}